\documentstyle[preprint,eqsecnum,aps]{revtex}
\begin{document}

\title{Midi-Superspace Models \\
of Canonical Quantum Gravity\cite{conference}
}
\author{C. G. Torre}
\address{Department of Physics\\
Utah State University\\
Logan, UT 84322-4415, USA
}

\date{June 1998}
\maketitle

\begin{abstract}
A midi-superspace model is a field theory obtained by
symmetry 
reduction of a parent gravitational theory. 
Such models have proven useful for exploring the 
classical and quantum dynamics of the gravitational
field.  I present 
3 recent classes of results pertinent to canonical
quantization of 
vacuum general relativity in the context of midi-superspace
models. 
(1) I give necessary and sufficient conditions such that a
given 
symmetry reduction can be performed at the level of the
Lagrangian or 
Hamiltonian. (2) I discuss the Hamiltonian formulation of 
models based upon cylindrical and toroidal symmetry. In 
particular, I explain how 
 these models can be identified with parametrized field
theories 
of wave maps, thus a natural strategy for canonical
quantization 
is available. (3) The quantization of a parametrized field
theory, 
such as the midi-superspace models considered in (2),
requires 
construction of a 
quantum field theory on a fixed (flat) spacetime that
allows for time 
evolution along arbitrary foliations of spacetime. I
discuss some 
recent results on the possibility of finding such a quantum
field 
theory.
\end{abstract}


\section{Introduction}
\label{sec: intro}

A time-honored strategy for extracting information from a 
field theory is to 
restrict attention to states of the system that possess
some degree of 
symmetry.  This strategy was applied to canonical quantum
gravity by 
DeWitt in 1967 \cite{DeWitt1967}.  He studied the canonical
quantization 
of spacetimes with matter that were
homogeneous and isotropic.  
This symmetry assumption turns 
an intractable problem in quantum field theory into a
straightforward 
quantum mechanical problem that can be solved completely
and 
illuminates some of the qualitative features (conceptual 
if not technical) of the full theory.   
A couple of years later, Misner dropped the isotropy
assumption and 
studied classical and quantum dynamics of homogeneous
spacetimes 
\cite{Misner1969}.  The 
field of quantum cosmology was born.

When using a description of gravitational dynamics based
upon metrics, 
it is reasonable to try to represent
the quantum states of the gravitational field 
 as functions of three-dimensional spatial geometry
\cite{DeWitt1967}.  
 This domain for 
the quantum gravitational state function has come to be
known as ``superspace''.  Naturally, if one restricts
attention to a small 
subspace of superspace, such as is done in quantum
cosmology, one 
naturally is dealing with a ``mini-superspace''. 
 The mini-superspace models 
of 
gravity consider spacetimes with so much symmetry that 
there are only a finite number 
of degrees of freedom left in the gravitational field.  A
few years after 
DeWitt and Misner began the mini-superspace/quantum
cosmology 
program, Kucha\v r went one step further
\cite{Kuchar1971}.  He considered the 
quantization of spacetimes admitting cylindrical 
symmetry.  This 
symmetry assumption was strong enough for him to make
progress in 
understanding canonical quantum gravity, but weak enough so
that the 
number of degrees of freedom in the gravitational field
were still 
infinite. Indeed, the symmetry assumptions made in
\cite{Kuchar1971} 
reduce the gravitational dynamics to that of Einstein-Rosen
waves.  
Borrowing from the vernacular of the fashion world, 
Kucha\v r proposed to call this model a ``midi-superspace''
model.  
Whether discussing 
mini-superspaces or midi-superspaces, one is  referring 
to symmetry reductions of 
the gravitational field.  Mini-superspaces lead to
mechanical 
models; the reduced field equations become ordinary
differential equations. 
Midi-superspaces define field theories; the reduced field 
equations  remain partial differential equations.

Since the pioneering work of DeWitt, Misner, and 
Kucha\v r, the amount of effort spent in studying classical 
and quantum properties of 
symmetry reductions of 
general relativity (and other field theories) has been 
enormous.  My purpose here is not to review this sizable
body of 
literature but rather to highlight some recent developments
in the 
area.  Of course, the choice of material has a strong
editorial bias.  
Most of the work that I will discuss was performed in 
collaboration with others.  In particular I have benefited
from 
working with Ian Anderson, Mark Fels, Joseph Romano, and
Madhavan 
Varadarajan on the topics discussed below.  Needless to
say, I take 
credit for the inevitable weaknesses in the presentation.

The following 3 topics will be discussed here. 

\medskip\noindent
{\bf (1)} {\sl The Symmetric Criticality Principle:} What
are necessary and 
sufficient conditions 
such that the canonical structure of a
field theory induces a canonical structure on a given
symmetry reduction 
of the theory?  This issue is clearly relevant when one is
studying 
canonical quantization of a mini- or midi-superspace, since
one needs 
to know what is the Hamiltonian, what are the constraints, 
what are the Poisson algebras of various 
functions, {\it etc.}  The issue boils down to the
question: When can one symmetry 
reduce a classical field theory via a symmetry reduction of
the 
Lagrangian?  In the mathematics literature this question
has 
appeared as the question of the validity of the
 ``symmetric criticality principle'' \cite{Palais1979}.

\medskip\noindent
{\bf (2)} {\sl Two Killing Vector Midi-Superspaces}: I will
consider some relatively recent results on the Hamiltonian 
structure of a popular class of midi-superspace models 
in which one assumes cylindrical symmetry or toroidal
symmetry.  
These are examples of ``2 Killing vector models''. The
results 
presented are generalizations of the work of Kucha\v r, and
lead to a nice scheme for canonical quantization.

\medskip\noindent
{\bf (3)} {\sl Quantization on Curved Surfaces}: 
The models considered in (2) allow one to turn the problem 
of finding a 
gauge-invariant canonical quantization of the 2 Killing
vector models 
into the problem of quantizing certain parametrized field
theories. 
 The problem of quantizing parametrized field theories 
is an interesting problem 
in its own right.  Dirac seems to be one of the first to
study this problem 
\cite{Dirac1964};
he called it the 
problem of {\it quantization on curved surfaces}.  
I will present some salient results in this area.

\section{The Symmetric Criticality Principle}
\label{sec: SCP}

Symmetry reduction of a classical field theory takes place
in 3 
steps.  First, one specifies a group action 
 with respect to which the fields are to be invariant. 
Second, 
one constructs the most general field admitting the chosen
group 
action as a symmetry.  This is the  {\it invariant
field}.  Normally, 
 the invariant 
field involves arbitrary functions of one or more
variables.  These are 
the {\it reduced fields}, which define the mini- or
midi-superspace. 
Third, one evaluates the field equations on 
the invariant fields thus obtaining the differential
equations (or 
perhaps algebraic equations) for the 
reduced fields.  These equations are the {\it reduced field 
equations} for the mini- or midi-superspace.  
All of us at one time or another have performed symmetry 
reduction in this way, {\it e.g.,} using time translation
and 
rotational symmetry to find the Schwarzschild solution of
the 
vacuum Einstein equations.  

Normally, the field equations can be derived from a
Lagrangian, in 
which case 
there is a very tempting shortcut one might try to obtain
the reduced field 
equations.  One can try substituting the invariant field
into the 
Lagrangian, thereby obtaining a {\it reduced Lagrangian}
for the reduced fields.  
One can then compute Euler-Lagrange equations from this
reduced 
Lagrangian and  obtain a set of reduced field equations. 
The only 
difficulty with this shortcut is that there is no 
guarantee  the reduced Lagrangian will yield the correct
reduced 
field equations!  This problem is particularly vexing when
studying 
quantization of mini- or midi-superspaces since one would
like to assume 
that restriction of the Lagrangian (or Hamiltonian) to the 
invariant fields yields the correct Lagrangian (or
Hamiltonian) 
 for the mini- or midi-superspace.

 A formal statement of the issue at hand is as 
follows.  Let ${\cal Q}$ be the space of metrics $g$ on a
manifold $M$.  
Let $S\colon {\cal Q}\to {\bf R}$ be an action functional
invariant 
under some transformation group $G\colon {\cal Q}\to {\cal
Q}$.  
Let $\hat{\cal Q}$ be the space of $G$-invariant metrics
$\hat g$ and let 
$\hat S\colon \hat{\cal Q} \to \hat{\cal Q}$ be the
restriction of 
$S$ to the invariant metrics. We want to know if the
critical points 
of $\hat S$ define critical points of $S$, that is,  
\begin{equation}
{\delta S\over \delta g}\Bigg|_{\hat {\cal Q}} = 0
\Longleftrightarrow {\delta \hat S\over \delta \hat g} = 0.
\end{equation}
Put simply: we want to 
know if symmetric critical points are critical symmetric
points.  If 
this is the case, following Palais, we say that the {\it
symmetric 
criticality principle} holds \cite{Palais1979}.  In
essence, if the symmetric criticality 
principle holds then one can expect to describe the reduced
field theory using 
the reduced Lagrangian.
Let us look at a couple of examples.

First we consider the spherical symmetry reduction of
vacuum general 
relativity.  We restrict attention to spacetimes admitting
a 
three dimensional isometry group $G$ that is isomorphic to
$SO(3)$ and
has orbits that are spacelike and diffeomorphic to
two-dimensional 
spheres.  It follows that there will exist (non-unique)
$G$-invariant functions, 
$t\in(-\infty,\infty)$ and $r\in(0,\infty)$, which can be
used as coordinates 
and
 such that the spacetime 
metric $g$ takes the form
\begin{equation}
g=A(r,t) dt\otimes dt + B(r,t) dt\otimes dr + C(r,t)
dr\otimes dr
+ D(r,t) d\Omega.
\label{so(3)metric}
\end{equation}
Here the group orbits are labeled by $t$ and $r$, 
$d\Omega$ is 
the standard metric on the unit 2-sphere, and $A$, $B$,
$C$, $D$ are 
arbitrary functions (aside from the conditions required to
keep the metric 
non-degenerate and to give it the correct signature).  We
obtain the 
conditions for the  
invariant metric (\ref{so(3)metric}) to define a
vacuum spacetime by requiring that this metric have
vanishing 
Einstein
tensor $G_{ab}$.  This requirement yields a system of 10
PDEs in 2 independent variables 
for the four functions $A$, $B$, $C$, $D$:
 \begin{equation}
G_{ab}[A,B,C,D] = 0,
\label{so(3)eom}
\end{equation}
 but it 
turns out that only 4 of the equations are independent. 
Thus we 
obtain 4 reduced field equations for 4 reduced fields, from
which one 
can obtain the Schwarzschild solution, {\it etc.}  Now we
can try 
our shortcut.  Substitute the metric (\ref{so(3)metric})
into the 
Einstein-Hilbert Lagrangian density, $L=\sqrt{g} R$, where
$R$ is the 
scalar curvature of the metric.  Up to an irrelevant factor 
coming from the area element on the unit 2-sphere, the
result is a Lagrangian 
density, $\hat L$, for the reduced fields.  The
Euler-Lagrange equations for the 
reduced fields obtained from $\hat L$ are a system of 4
PDEs in 2 
independent variables:
\begin{equation}
{\delta \hat S\over \delta A}
={\delta \hat S\over \delta B}
={\delta \hat S\over \delta C}
={\delta \hat S\over \delta D}=0.
\label{so(3)el}
\end{equation}
The equations (\ref{so(3)el}) can be shown to be equivalent
to the equations 
(\ref{so(3)eom}), so the reduced Lagrangian $\hat L$
correctly describes the spherically symmetric vacuum
spacetimes.  
Thus the symmetric criticality principle holds for
spherically 
symmetric reductions of the vacuum Einstein equations.  

The symmetric criticality principle for the spherical
symmetry reduction 
of general relativity 
was used by Weyl to derive the Schwarzschild solution 
\cite{Weyl1952}. The principle is also being used in recent
approaches 
to canonical quantization of midi-superspace models of 
spherically symmetric spacetimes (see {\it e.g.},
\cite{Kuchar94}).

As another example, let us consider homogeneous solutions
to the 
vacuum Einstein equations.  These mini-superspace
cosmological models 
are obtained by restricting attention to spacetimes that
admit a 
three-dimensional group $G$ of isometries with orbits
$\Sigma$ that are leaves of 
a co-dimension 1 spacelike foliation of spacetime. We label
the 
homogeneous hypersurfaces by $t$. The invariant metrics
take the 
form
\begin{equation}
g = \alpha(t) dt\otimes dt + \beta_{i}(t) dt\otimes
\omega^{i}
+\gamma_{ij}(t) \omega^{i}\otimes \omega^{j},
\label{homometric}
\end{equation}
where $\alpha$, $\beta_{i}$, and $\gamma_{ij}=\gamma_{ji}$
are arbitrary functions of $t$ 
(modulo non-degeneracy and signature of the metric), and
$\omega^{i}$, $i=1,2,3$, 
are a basis of $G$-invariant 1-forms on 
$\Sigma$.  The equations of motion for homogeneous vacuum
metrics are 
a system of 10 ODEs for the 10 functions  
$\alpha$, $\beta_{i}$ and $\gamma_{ij}$ obtained by 
demanding that the metric (\ref{homometric}) have vanishing
Einstein 
tensor:
\begin{equation}
G_{ab}[\alpha,\beta,\gamma]=0.
\label{homoeom}
\end{equation} 
Thus we obtain 10 reduced equations of motion for the 10
reduced fields.
As before, we can try to obtain the reduced equations of
motion from the 
reduced Lagrangian $\hat L$.  Substituting
(\ref{homometric}) into the 
Einstein-Hilbert Lagrangian and computing the
Euler-Lagrange 
equations for the reduced fields $\alpha$, $\beta_{i}$ and
$\gamma_{ij}$ we 
again obtain 10 ODEs:
\begin{equation}
{\delta \hat S\over\delta \alpha}={\delta \hat S\over\delta
\beta_{i}}
={\delta \hat S\over\delta \gamma_{ij}}=0.
\label{homoel}
\end{equation}
The equations (\ref{homoel}) are  equivalent 
to (\ref{homoeom}) only if the structure constants
$C_{ab}{}^{c}$ of 
$G$ satisfy
\begin{equation}
C_{ab}{}^{b}=0.
\label{typea}
\end{equation}
Homogeneity groups satisfying (\ref{typea}) have been 
given the picturesque name {\it class A}. 
(A more descriptive term for groups satisfying
(\ref{typea}) 
is {\it unimodular} since such groups admit a bi-invariant
volume form.)  If 
$G$ does not satisfy (\ref{typea}) it is, of course, called
{\it class 
B}.  
We see that the class A models obey the symmetric
criticality 
principle but that the class B models do not.  Thus the 
Einstein-Hilbert 
variational principle for the Einstein equations fails to
induce a 
variational principle for the class B homogeneous
cosmological models.
This annoying feature of homogeneous class B
mini-superspaces has been known 
for a long time (see {\it e.g.}, 
\cite{Mac79} and \cite{Shepley98} for a discussion).  
Needless to say, canonical quantization of mini-superspaces
describing homogeneous cosmologies has been studied only
for the class A models.

Experience with examples such as described above suggests
that 
the validity of the 
reduced Lagrangian (or Hamiltonian) for a given mini- or 
midi-superspace has to be checked on a theory-by-theory and
group-by-group 
basis.  Fortunately, it is possible to give general
conditions 
that are necessary and sufficient for the symmetric
criticality 
principle to be valid for any field theory. Palais gives
results 
along these lines in the context of $G$-invariant functions
on Banach 
manifolds \cite{Palais1979}.  
It is possible to give a somewhat more detailed set 
of conditions by specializing  to local Lagrangian field
theories 
\cite{AFT98}.  Here I would like to give an informal
statement 
of the main result from \cite{AFT98} on symmetric
criticality 
for local gravitational field 
theories. To state this 
result we need the following data.  The symmetry group
being used for 
reduction is $G$. The 
orbits of $G$ in spacetime $M$ have dimension $q$. 
The isotropy (or stabilizer) 
group of a point $x\in M$ 
is $H_{x}\subset G$. It is assumed that the isotropy group
of any 
given point 
is a subgroup of the Lorentz group since this is a
necessary and 
sufficient condition for the (local) existence of a
$G$-invariant metric 
\cite{AFT98}. 
The vector space of
symmetric rank-2 tensors at a 
point $x\in M$ is denoted by $V_{x}$. The vector space of 
$H_{x}$-invariant symmetric rank-2 tensors at a 
point $x$ of spacetime is denoted by $V^{H}_{x}$. The
annihilator of 
$V^{H}_{x}$ 
(linear functions on $V$ that vanish on $V^{H}$) is denoted
by 
$(V^{H}_{x})_{0}$.  Finally, the Lie algebra cohomology of
$G$ relative to 
$H$ at degree $q$ is denoted by ${\cal H}^{q}(G,H)$.  This
is the 
space of $H$-invariant closed {\it modulo} exact $q$-forms
on the group 
manifold for $G$.

\proclaim Theorem.
The principle of symmetric criticality is valid for any
metric field 
theory derivable from a local Lagrangian density if and
only if the 
following 2 conditions are satisfied at 
each point $x$ in the 
region of spacetime under consideration.

\bigskip
{\bf (1)} ${\cal H}^{q}(G,H_{x})\neq 0$.

\bigskip
{\bf (2)} $V^{H}_{x}\cap (V^{H}_{x})_{0}=0$.

\bigskip
 The appearance of these 
2 conditions can be understood from the 2 possible ways
that 
symmetric criticality can fail. Recall that the first
variation of an 
action functional is the sum of (i) a boundary term coming
from an 
integration by parts, and (ii) a volume term in which the
Euler-Lagrange 
expression occurs.  The reduced Euler-Lagrange 
equations fail to 
define  solutions of the original Euler-Lagrange equations 
if (a) the boundary term of the original variational 
expression fails to reduce to the boundary term for the
reduced 
Lagrangian and/or
(b) any (non-trivial) field equations coming from the
volume term of 
the original 
Lagrangian disappear in the volume 
term for the reduced Lagrangian.  If (a) occurs, then the
reduced 
Euler-Lagrange equations pick up additional terms that 
render the Euler-Lagrange equations incorrect. If (b)
occurs, then there will be non-trivial 
field equations that simply do not arise from the reduced
Lagrangian. 
Condition (1) in the theorem, 
which is equivalent to requiring that the group orbits
admit a 
bi-$G$-invariant volume form, 
is necessary and sufficient to guarantee that (a) will not
occur.  
Condition (2) is necessary and sufficient to prevent (b).

\bigskip\noindent
{\it Remarks:}

\medskip\noindent
$\bullet$
When I gave this talk I left out the 
second condition (2) in the statement of the theorem. At
the time, we 
thought that the existence of a $G$-invariant metric 
would prevent  problem (b) from arising. Problem (b) {\it
is} absent for 
field theories of a
Riemannian metric, and condition (2) is not needed in that
setting. 
But in general, and in particular for field theories of a 
Lorentzian metric, problem (b) can 
arise and condition (2) is needed.

\medskip\noindent 
$\bullet$ 
The conditions needed for the validity of the symmetric
criticality 
principle may seem a little arcane, but they are in fact
quite easy to 
check using elementary linear algebra and some manipulation
of 
structure constants for the symmetry group.

\medskip \noindent
$\bullet$ 
The theorem above gives conditions for validity of
reduction of  
{\it any}  Lagrangian.  It is possible for the 2 
conditions to fail for a given Lagrangian and still have
symmetric 
criticality holding {\it for that particular Lagrangian}. 
The 
utility of the theorem is that it allows you to take a
given symmetry 
reduction and check {\it a priori} whether one can
correctly 
symmetry reduce at the level of the Lagrangian,
irrespective 
of the choice of Lagrangian.  Perhaps I should also point
out that 
even if the symmetric criticality principle fails and 
the reduced Lagrangian does not correctly describe the 
reduced field equations, this does not mean that the
reduced field 
equations do not admit a variational principle of some
other type.

\medskip\noindent
$\bullet$ 
It can be shown that the conditions of the theorem are
satisfied if 
$G$ is compact. In particular, it is always safe to reduce 
 the Lagrangian via spherical symmetry (a fact often taken
for granted 
 in the physics literature!). If the group 
action is free, that is, has no (non-trivial) isotropy
subgroups, then condition (2) 
in the theorem is trivially satisfied and condition (1)
reduces to the 
statement that the Lie group is unimodular (\ref{typea}). 
Thus we 
recover the results of our two examples given above, and we
see that 
these results are not specific to the Einstein field
equations.

\medskip\noindent
$\bullet$ 
While the theorem above is stated in the context of a local
field 
theory of a metric, there is a straightforward
generalization of the 
theorem to essentially any type of local field theory 
\cite{Palais1979,AFT98}.

\section{Two Killing Vector Midi-Superspaces}
\label{sec: 2KV}

Let us now focus on a particular class of midi-superspace
models 
obtained by assuming the existence of 2 commuting Killing
vector 
fields for the spacetimes of interest.  These models have
an 
infinite number of degrees of freedom and are equivalent to 
field theories in  two-dimensions.  Thus these are among
the simplest of the 
midi-superspaces.  We shall present the Hamiltonian
formulation of 
these models and indicate that they are mathematically
equivalent to 
parametrized field theories. Thus their quantization can be
viewed as an 
instance of Dirac's problem of quantization on curved
surfaces.

We will study two of the ``2 Killing vector models'',
namely, a 
cylindrically symmetric model and a toroidally symmetric
model. The 
former is a generalization of the Einstein-Rosen wave model
of Kucha\v 
r \cite{Kuchar1971}. The latter is the ``Gowdy model''
\cite{Gowdy74}.  
We begin by defining the 
spacetime manifold and symmetry group.

\bigskip\noindent
{\it Cylindrical Symmetry}

Here the spacetime manifold is $M={\bf R}\times {\bf R}^{3}$
 with  cylindrical coordinates $(t,x,\phi,z)$, where
\begin{equation}
t\in(-\infty,\infty),\quad
x\in(0,\infty),\quad
\phi\in(0,2\pi),\quad
z\in(-\infty,\infty).
\end{equation}
The symmetry group is generated by a translation, a
rotation, and a 
discrete $Z_{2}$. The translation 
and rotation are generated by the vector fields 
$({\partial\over\partial z},{\partial\over\partial \phi})$,
\begin{equation}
\phi\to\phi+{\rm constant}\ modulo\ 2\pi,
\quad z\to z +{\rm constant},
\end{equation} 
 and the 
discrete transformation is
\begin{equation}
(t,x,\phi,z)\longrightarrow (t,x,2\pi-\phi,-z).
\end{equation}

\bigskip\noindent
{\it Toroidal Symmetry}

Here the spacetime manifold is $M={\bf R}\times {\bf
T}^{3}$ 
with  coordinates $(t,x,y,z)$, where
\begin{equation}
t\in(-\infty,\infty),\quad
x,y,z\in (0,2\pi).
\end{equation}
Note that each of $(x,y,z)$ are angular coordinates on a
3-torus.
The symmetry group is generated by  two rotations on the
torus and a 
$Z_{2}$ again. The rotations are generated by the vector
fields 
$({\partial\over\partial y},{\partial\over\partial z})$,
\begin{equation}
y\to y+{\rm constant}\ modulo\ 2\pi,
\quad z\to z +{\rm constant} \ modulo\ 2\pi,
\end{equation}  
 and the 
discrete transformation is
\begin{equation}
(t,x,y,z)\longrightarrow (t,x,2\pi-y,2\pi-z).
\end{equation}

In each case the discrete symmetry is designed to force 
the orthogonal distribution associated with the Killing
vector fields 
to be integrable. In the usual terminology, we are
considering spacetimes 
admitting two 
commuting Killing vector fields generating an
``orthogonally transitive'' 
group action.

It is straightforward to find the general form of the
metrics 
admitting the toroidal and cylindrical symmetry groups.  
We present their line 
elements in the coordinates described above \cite{RT96}.

\bigskip\noindent
{\it Cylindrical Symmetry}

\begin{equation}
ds^{2}=\left[-(N^{\perp})^{2}+e^{\gamma-\psi}(N^{x})^{2}
\right]dt^{2}
+2e^{\gamma-\psi}N^{x}dtdx + e^{\gamma-\psi}dx^{2}
+\Phi^{2}e^{-\psi}d\phi^{2} +e^{\psi}(dz + \tilde\psi
d\phi)^{2}.
\end{equation}
Each of the 6 fields entering into the components of the
metric are 
functions of $t$ and $x$ only.  We assume that $\Phi>0$,
that the spacetime gradient 
of  $\Phi$ is everywhere spacelike, and that $N^{\perp}>0$. 
The reduced fields
$(N^{\perp},N^{x},\gamma,\Phi,\psi,\tilde\psi)$ 
are otherwise unrestricted. The variables $N^{\perp}$ and
$N^{x}$ are 
the lapse and shift for a symmetry compatible foliation
(see \cite{MTW} for a 
discussion of the 3+1 formalism). 
The Einstein equations are 6 non-linear 
PDEs for the six reduced fields. The ``true degrees of
freedom'' of 
the model can be identified with the fields
$(\psi,\tilde\psi)$ \cite{RT96}.

\bigskip\noindent
{\it Toroidal Symmetry}

\begin{equation}
ds^{2}=\left[-(N^{\perp})^{2}+e^{\gamma-\psi}(N^{x})^{2}
\right]dt^{2}
+2e^{\gamma-\psi}N^{x}dtdx + e^{\gamma-\psi}dx^{2}
+\Phi^{2}e^{-\psi}dy^{2} +e^{\psi}(dz + \tilde\psi dy)^{2}.
\end{equation}
Each of the 6 fields entering into the components of the
metric are 
functions of $t$ and $x$ only.  We assume that $\Phi>0$,
that the spacetime gradient 
of  $\Phi$ is everywhere timelike, and that $N^{\perp}>0$. 
The reduced fields
$(N^{\perp},N^{x},\gamma,\Phi,\psi,\tilde\psi)$ 
are otherwise unrestricted.  The variables $N^{\perp}$ and
$N^{x}$ are 
the lapse and shift for a symmetry compatible foliation.
The Einstein equations are 6 non-linear 
PDEs for the six reduced fields. The ``true degrees of
freedom'' of 
the model can be identified with the fields
$(\psi,\tilde\psi)$ 
subject to a single ``zero momentum'' constraint and a
single ``point 
particle'' degree of freedom \cite{RT96}.

Let us check that the 2 Killing vector midi-superspace
models acquire a 
Hamiltonian structure from the Hamiltonian structure of the
full 
theory (see \cite{MTW,HRT,Teitelboim80,AA} for a discussion
of Hamiltonian gravity). 
This will follow if the symmetric criticality principle is
valid for 
the cylindrical and toroidal symmetry reductions. 
 We therefore check whether the two conditions from the 
theorem of the last 
section are satisfied for these symmetry reductions. 
That the first condition is satisfied follows 
from the fact that the group orbits admit an invariant
volume form, 
{\it e.g.}, $d\phi \wedge dz$ (cylindrical symmetry) or
$dy\wedge 
dz$ (toroidal symmetry). That the second condition is
satisfied 
follows from the fact that the representation of the
($Z_{2}$) linear isotropy 
group at any point of 
the spacetime manifold
 is fully reducible.  This second condition can also 
 be checked by direct computation of the spaces $V^{H}$ and 
$V^{H}_{0}$. We conclude that the symmetric criticality
principle 
applies to cylindrical or toroidal symmetry reductions of
any metric 
field theory.  Therefore, it is permissible to derive the
Hamiltonian 
formulation of these models by restricting the full
Hamiltonian 
formulation to the chosen midi-superspace.  This justifies
the usual 
procedure, often found in the literature, in which the
general 
form of the invariant metric is substituted into the ADM
action to obtain the ADM 
action for the reduced theory.

It is possible to give a more or less unified treatment of
the Hamiltonian 
formulation of the models we are considering. The most
succinct way 
to do this is to display the phase space (or ``ADM'')
action functional:
\begin{equation}
S=\int_{\hat M}\left(\pi_{\Phi}\dot \Phi
+\pi_{\gamma}\dot\gamma +\pi_{\psi}\dot\psi 
+\pi_{\tilde\psi}\dot{\tilde\psi}
-N^{\perp}{\cal H}_{\perp} - N^{x}{\cal H}_{x}\right) +\
{\rm 
boundary\ 
term}.
\label{2kvaction}
\end{equation}
The structure of (\ref{2kvaction}) is as follows. The
integral is over 
the space of orbits, $\hat M$ (with coordinates $(t,x)$), 
of the Killing vector fields.  Thus 
$\hat M={\bf R}\times {\bf R}^{+}$ in the cylindrical
symmetry case, 
and $\hat M={\bf R}\times S^{1}$ in the toroidal symmetry
case.  The 
critical points of this action functional define vacuum
spacetimes 
with the prescribed symmetries. To find these critical
points the 
action is to be varied with respect to 
 the midi-superspace fields ($\gamma$, $\Phi$, $\psi$,
$\tilde 
\psi$), their conjugate momenta ($\pi_{\gamma}$,
$\pi_{\Phi}$, 
$\pi_{\psi}$,
$\pi_{\tilde\psi}$), along with the lapse and shift
($N^{\perp}$, 
$N^{x}$). The latter two variations lead to the (symmetry
reduced) 
Hamiltonian and momentum constraints:
\begin{equation}
{\cal H}_{\perp}:=
e^{(\psi-\gamma)/2}\left[
-\pi_{\gamma}\pi_{\Phi} + 2\Phi^{\prime\prime}
-\Phi^{\prime}\gamma^{\prime}
+{1\over2}(\Phi\psi^{\prime 2} + \Phi^{-1}\pi_{\psi}^{2})
+{1\over2}(\Phi e^{-2\psi}\pi_{\tilde\psi}^{2}
+\Phi^{-1}e^{2\psi}\tilde{\psi^{\prime 2}})\right]=0,
\label{scalar}
\end{equation}

\begin{equation}
{\cal H}_{x}:=-2\pi_{\gamma}^{\prime}
+\pi_{\gamma}\gamma^{\prime} 
+\pi_{\Phi}\Phi^{\prime}+\pi_{\psi}\psi^{\prime}
+\pi_{\tilde\psi}\tilde\psi^{\prime}=0,
\label{vector}
\end{equation}
where a prime indicates a derivative with respect to the
spatial 
coordinate $x$.  The boundary term indicated in
(\ref{2kvaction}) 
only arises in the cylindrically 
symmetric case and is needed to render the action and
Hamiltonian 
differentiable with appropriate boundary conditions. I 
refer you 
to \cite{RT96} for details on this point.

The appearance of a pair of constraints on the phase space
of the 
midi-superspace models reflects the presence of a 
gauge symmetry of the models with respect to a
two-dimensional 
diffeomorphism group.  This is the group of diffeomorphisms
of the 
symmetry-reduced spacetime manifold, {\it i.e.}, the
space of orbits $\hat M$ of the Killing vector fields. 
Indeed, one can check 
by direct computation that the Poisson algebra of the
constraint 
functions $({\cal H}_{\perp},{\cal H}_{x})$ is the Dirac
algebra of 
hypersurface (really,  curve) deformations in a
two-dimensional 
spacetime, which is the Hamiltonian expression of general
covariance 
\cite{Teitelboim80}. As in the full theory of gravity, the
Hamiltonian is (up 
to surface terms) built from the constraint functions, so
that the 
constraint functions generate (almost all) of the dynamics.
It is 
the existence of a Hamiltonian 
and momentum constraint that makes
the 2 Killing vector midi-superspaces such 
excellent models of canonical quantum gravity.    

Unfortunately, the constraint functions  ${\cal H}_{\perp}$
and ${\cal 
H}_{x}$, as they stand, are still rather intractable from
the point of 
view of quantization. A straightforward approach {\it a la}
Dirac 
\cite{Dirac1964}
(see also \cite{DeWitt1967} and \cite{Kuchar1971}) would go
as follows. 
Build the state space of the quantum 
theory as a space of 
functionals of the midi-superspace variables ($\gamma$,
$\Phi$, $\psi$, $\tilde 
\psi$) modulo spatial ($x$) diffeomorphisms.  By taking the
quotient 
with respect to the action 
of spatial diffeomorphisms we take into account the quantum 
form of the momentum constraint (\ref{vector}).  
What is left is the midi-superspace 
version of the Wheeler-DeWitt equation, in which one
imposes the 
quantum form of the constraint (\ref{scalar}), say, by
trying to represent 
the 
midi-superspace variables as multiplication operators,
representing their 
conjugate momenta as functional derivative operators, and
then 
demanding that this quantization of ${\cal H}_{\perp}$
annihilate 
physical states.  No one seems to have made any progress
using 
this most direct of approaches. One might say that the 2
Killing vector 
midi-superspaces 
model the situation in vacuum geometrodynamics all too
well.  To my 
knowledge, progress on these models has been made using 3
alternative 
approaches.  

First, one can translate the midi-superspace model into 
the phase space description based upon Ashtekar variables
\cite{AA}.  As in 
the full theory of gravity, this leads to a significantly
different 
set of strategies for the quantization of the 
constraints.  This approach was initiated in \cite{smolin}
and some preliminary 
results obtained.  Neville has developed this approach in
some detail 
for the plane wave midi-superspaces \cite{neville}.  I must
direct 
you to these references for details.

Second, one can simply eliminate the diffeomorphism 
invariance (at the classical level) using coordinate
conditions. 
In this approach, the 
constraints are solved classically and the issues of
general 
covariance, constraint quantization, {\it etc.}, are
eliminated from 
consideration. There is a nice gauge fixing for the models
being 
considered here based 
upon the Einstein-Rosen (cylindrical symmetry) or Gowdy
(toroidal 
symmetry) coordinates.  If the dynamics of the theory are
restricted 
to foliations of spacetime adapted to these coordinate
systems, then 
the dynamics of the theory are (modulo a few subtleties,
see below) 
mathematically equivalent to that of a symmetry reduction
of a 
wave map \cite{note2} 
from a flat three-dimensional spacetime to a 
two-dimensional Riemannian manifold of constant negative
curvature. 
The wave map is provided by the fields $\psi$ and
$\tilde\psi$. In 
this interpretation of the symmetry reduced, gauge fixed
theory, 
the space of orbits $\hat M$ is 
viewed as a symmetry reduction of a flat three dimensional
spacetime 
by a one dimensional group.
By completely fixing the gauge, one can thus turn the
quantization problem to the study of 
the quantum theory of wave 
maps on a flat spacetime.  This point of view is the one
taken in 
\cite{KS} and \cite{Ashtekar96}.  
In 
particular, much progress has been made for the case where
one 
assumes that the Killing vector fields are hypersurface
orthogonal.  
This removes one ``polarization'' from the gravitational
field and 
reduces the wave map to a single free scalar field, which
is the 
Einstein-Rosen wave amplitude in the cylindrically
symmetric case.  
Because the model has been reduced, 
in effect, to a 
free field theory, one can say quite a bit about the
quantum theory.
It is gratifying to be able to 
extract quantum information about spacetime geometry in
this 
field-theoretic setting. 

While the gauge-fixed quantum theory of the midi-superspace
models has 
shed new light on possible physical properties of quantum
geometry, the models 
fail to help us understand the full theory in one important
respect. 
In the full theory one 
aspires to formulate the quantization in such a way as to
preserve general covariance.  This means one keeps the
constraints in 
the theory and quantizes the constrained theory {\it a 
la} Dirac \cite{Dirac1964}.  Presumably, the results
obtained for the 
fully gauge fixed models arise as a specialization of the
putative 
gauge invariant quantum theory.  Without the ability to
appeal to a 
gauge invariant formulation, it is not clear how to relate
the results 
obtained via different gauge fixing methods. Indeed, it is
hard 
to be sure that the gauge fixed theory has been quantized
in a manner 
consistent with general covariance.

A third 
approach is possible, which still takes advantage of the
wave map nature 
of the true degrees of freedom, but which preserves general 
covariance so as to provide a viable model of Dirac
constraint 
quantization of the full theory.  This approach is a
generalization of that used by Kucha\v 
r, and is based upon a canonical transformation that
identifies the 
midi-superspace model with a {\it parametrized field
theory} of 
wave maps. I would now like to describe this approach in
more detail.

A parametrized field theory is a field theoretic
generalization of a familiar 
construction from mechanics (see, e.g., \cite{Lanczos}) 
in which one introduces a 
new, arbitrary time 
parameter $\tau$ into the system and formally treats the
true 
time $t$
as a new degree of freedom which evolves in the new time
parameter. 
Of course, time is not a degree of freedom: the time $t$ is
an 
arbitrary function of the parameter time, $t=t(\tau)$. 
This fact manifests itself in the appearance of a
constraint in 
the Hamiltonian formulation that identifies the momentum
conjugate to 
time with (minus) the canonical
energy of the system. One can interpret the appearance of
the 
constraint as reflecting a diffeomorphism gauge symmetry of
the problem 
associated 
with the arbitrariness of the new time parameter $\tau$. 
The generalization 
of this formalism to field theory is reasonably
straightforward 
\cite{Dirac1964,HRT,IK}. In field theory, an instant of
time 
is a Cauchy hypersurface. 
Given a field theory on a fixed spacetime background, one
can express 
the theory in terms of an arbitrary foliation of the
background and 
treat instants of time (Cauchy surfaces) as new dynamical
variables.  
A Cauchy surface can be determined by giving its embedding
in the 
given spacetime. For a four-dimensional spacetime, this
means that one 
must specify 4 functions of 3 variables. These 4 functions,
along with 
their canonical momenta, are added to the phase space 
of the field theory to obtain 
the parametrized field theory.  As in the mechanical case,
adding 
time to the phase space of the theory also adds constraints
to the 
phase space. Four functions of the canonical variables must
vanish. These 
constraints take the following form:
\begin{equation}
{\cal C}_{\alpha}(x) := P_{\alpha}(x) +
h_{\alpha}[\phi,\pi,T](x)=0.
\label{pft}
\end{equation}
The notation used in (\ref{pft}) is as follows. Coordinates
on 
spacetime are denoted by ${\rm T}^{\alpha}$.  An embedding
of a 
Cauchy surface $\Sigma$ is given parametrically via
\begin{equation}
{\rm T}^{\alpha}=T^{\alpha}(x),
\end{equation}
where $x^{i}$ are coordinates on $\Sigma$.  The functions 
$T^{\alpha}(x)$ are dynamical variables in the parametrized
field theory.  Their 
conjugate momenta are denoted by $P_{\alpha}(x)$. The true
degrees of 
freedom of the theory are represented by the 
canonical variables $(\phi,\pi)$. The quantities 
$h_{\alpha}[\phi,\pi,T](x)$ are the flux of energy-momentum
at the 
spacetime point $T^{\alpha}(x)$ associated with the Cauchy
surface 
embedded by $T^{\alpha}$. In formulas:
\begin{equation}
h_{\alpha}=\sqrt{\gamma}n^{\beta}\Theta_{\alpha\beta},
\end{equation}
where $\Theta_{\alpha\beta}$ is the energy momentum tensor
for the 
fields $(\phi,\pi)$, $n^{\alpha}$ is the timelike unit
normal to 
the hypersurface defined by $T^{\alpha}(x)$ and
$\gamma_{ij}$ is the 
induced metric on that hypersurface.  

As you can see, the constraint 
(\ref{pft}) 
identifies the variable conjugate to time (space) with
minus the 
energy (momentum) density of the true degrees of freedom,
in complete 
analogy with the mechanical version of the parametrized
system.
The presence of the constraints in the parametrized field
theory 
reflect the existence of a four-dimensional diffeomorphism
gauge 
symmetry for the theory.  These constraints are ``first
class''; 
in fact the Poisson algebra of the constraints (\ref{pft})
is Abelian.  This 
allows one to make a direct connection between the
canonical 
transformations generated by the constraint functions and
the action 
of spacetime diffeomorphisms on the parametrized field
theory \cite{IK}. The 
Dirac algebra of hypersurface deformations is obtained by
taking projections 
of the constraints (\ref{pft}) normally and tangentially to
the Cauchy 
surfaces and computing their Poisson algebra.

To get a diffeomorphism symmetry for a field theory on a
fixed 
background we have to 
add variables to the theory, that is, we have to
``parametrize'' the 
theory to make it generally covariant.  Of course, in
general relativity these 
variables are, in some sense, already there, and one often
calls 
general relativity an ``already parametrized field
theory''.  Unlike 
the case with an  already parametrized theory such as
general 
relativity, when parameterizing a 
field theory on a fixed background spacetime the
constraints that 
appear have a very simple structure.  They indicate
explicitly that 4 canonical 
pairs are not truly dynamical.  Moreover, the constraint
functions
 generate the dynamical evolution of the 
true degrees of freedom as one deforms the embedding upon
which the 
degrees of freedom 
are being measured.  This leads to a natural approach to
Dirac 
constraint quantization of a parametrized field theory. The
embedding 
momenta are defined as variational derivative operators
with 
respect to the embeddings $X^{\alpha}(x)$, and one must
define the 
quantum energy-momentum flux as a self-adjoint operator on
a Hilbert 
space of states for the true degrees of freedom.  The
quantum 
constraints then, at least formally, constitute a
functional Schr\"odinger (or 
Tomonaga-Schwinger) equation
\begin{equation}
\left(
{1\over i}{\delta\over\delta T^{\alpha}} + h_{\alpha}\right)
|\Psi> = 0,
\label{TSeqn}
\end{equation}
 which defines the dynamical evolution of 
the state vector $|\Psi>$ along an arbitrary foliation of
spacetime.

We have remarked that by gauge fixing the midi-superspace
models one 
ends up with a (symmetry reduction) of a field theory of
wave 
maps on a flat spacetime.  Such a field theory can be made
generally covariant by the 
parametrization process sketched above, and it is natural
to ask if 
this parametrized field theory has anything to do with the
``already 
parametrized'' midi-superspace model.  The answer is
affirmative. 
By generalizing Kucha\v r's 
treatment of Einstein-Rosen waves, it is possible to
identify the 
cylindrical and/or toroidal midi-superspace models with
parametrized 
field theories  of  a one-dimensional symmetry reduction of
wave 
maps 
from a flat three-dimensional spacetime to a
two-dimensional 
Riemannian manifold of constant 
negative curvature.  More precisely, given a slight
modification of the 
phase spaces for the midi-superspace models being discussed
(see 
the remark
below), there exist canonical transformations identifying
these models 
with parametrized field theories of wave maps \cite{RT96}.
Thus the 
constraints (\ref{scalar}) and (\ref{vector}) 
of the toroidal or cylindrical midi-superspaces can be 
expressed  in the parametrized field theory form 
(\ref{pft}), and a clear strategy for implementing these 
constraints in quantum theory is thus available.  This
strategy for 
quantization was 
explored by Kucha\v r, albeit at a rather formal level, for
Einstein-Rosen waves 
in \cite{Kuchar1971}.  Teitelboim proposed that this same
strategy 
could be used to formulate the canonical quantization of
the full 
theory \cite{teitelboim75}.  Limitations on this approach
are 
discussed in \cite{CGT92}.

To use these results to implement Dirac constraint
quantization of the 
midi-superspaces being discussed here we must be able to
construct a 
quantum field theory on a fixed (indeed, flat) 
spacetime that allows for dynamical 
evolution along 
arbitrary foliations of spacetime.  Only relatively 
recently have results on the 
possibility of doing this become available.  We will take
up this 
issue in the next section.

\bigskip\noindent
{\it Remark:}

To make a rigorous identification of the midi-superspace
models with 
a parametrized field theory, one must extend slightly 
the definition of the phase space for the midi-superspace
models. The 
reason for this is that one is aspiring to use the
intrinsic and 
extrinsic geometry of a hypersurface (the geometric
interpretation of 
the gravitational phase space variables) to define how that
hypersurface 
is embedded in spacetime.  As it happens, the geometry of a 
hypersurface is not quite adequate to determine its
embedding into 
spacetime. One must add a single variable to the phase
space of the 
cylindrical symmetry model, and a pair of variables must be
added to 
the toroidal symmetry model to allow for the identification
with a 
parametrized field theory of wave maps.  In the toroidal
symmetry 
case there is also a a single extra constraint that
augments the usual 
constraints (\ref{pft}) of the parametrized field theory.
The meaning of this 
constraint is that the total momentum for the wave maps
must vanish. 
For details on all 
this, see \cite{RT96}.

\section{Quantization on Curved Surfaces}
\label{sec:pft}

Let us summarize the discussion of the last section. 
Cylindrically symmetric and toroidally symmetric
midi-superspace 
models of vacuum gravity are mathematically 
described by (a symmetry reduction of) 
parametrized field theories of wave maps on a flat 
spacetime.  This allows us to turn the problem of Dirac
constraint 
quantization of the 
midi-superspace models to that of finding a quantization of
fields on 
a fixed background spacetime 
that allows for dynamical evolution along an arbitrary
foliation of 
spacetime by Cauchy surfaces.  

Remarkably, in the same 1964 monograph \cite{Dirac1964} 
in which Dirac described his methods for handling
constrained 
Hamiltonian systems, he also considered the 
problem of canonically quantizing a field theory on
Minkowski 
spacetime such that one could consistently evolve the state
of the 
system from one arbitrary spacelike hypersurface to
another. He called 
this the problem of ``quantization on curved surfaces''. 
He 
formulated the problem in terms of the associated
parametrized field 
theory and indicated that, in general, one could expect
difficulties with 
consistency due to factor ordering problems in the quantum 
constraints (\ref{TSeqn}).

As far as I can tell, it took over 20 years before an
example of 
quantization on 
curved surfaces was worked out in any detail. I have in
mind the work of 
Kucha\v r in \cite{KK89}, 
followed by work of Varadarajan and myself \cite{TV98},
which considers 
the problem in the context of a free scalar field theory on
a flat 
two-dimensional background.  The problem is already rather 
interesting (and has only been explored)
for free fields, so let me try to describe what is going on
in that 
case only.

Let us first consider classical time evolution for a free
field 
$\varphi$ on a globally hyperbolic spacetime $(M,g)$.  
The evolution is determined by the field equations which 
we write as
\begin{equation}
\Delta(\varphi) = 0,
\label{eom}
\end{equation}
where $\Delta$ is a linear differential operator, {\it
e.g.}, 
$\varphi$ is a scalar field and
$\Delta=\nabla_{a}\nabla^{a}-m^{2}$ 
is the Klein-Gordon operator.  Denote by ${\cal S}$ the
space of 
solutions to (\ref{eom}) with appropriate boundary
conditions, say, 
compactly supported Cauchy data on any Cauchy surface
$\Sigma$.    
Denote by 
$\Gamma$ the space of Cauchy data for (\ref{eom}). Because
the Cauchy problem 
is well-posed, there is for each Cauchy surface $\Sigma$ an
isomorphism 
$e\colon\Gamma\to{\cal S}$ which takes Cauchy data on that
surface 
and yields the unique solution to (\ref{eom}) with that
data. 
The inverse, $e^{-1}\colon {\cal S}\to \Gamma$, takes a 
solution and yields its Cauchy data on $\Sigma$.
Assuming that the free field has no 
gauge symmetries, there will exist a (weak) symplectic form
on the 
space of solutions, that is, a 
skew-bilinear, non-degenerate map 
$\Omega\colon{\cal S}\times{\cal S}\to {\bf R}$. Given a
Cauchy 
surface, the symplectic form 
on ${\cal S}$ can be pulled back to $\Gamma$ using $e$ to
define a 
symplectic form $\omega\colon\Gamma\times\Gamma\to {\bf R}$:
\begin{equation}
\omega=e^{*}\Omega.
\end{equation}
It can be shown that $\omega$ is independent of the choice
of Cauchy 
surface used to define $e$. 
Either of the symplectic vector spaces $({\cal S},\Omega)$
or $(\Gamma,\omega)$ 
can be viewed as the 
phase space for the free field theory.

In order to discuss dynamics, we consider evolution 
from some initial time to some final 
time. Let us therefore fix an initial Cauchy hypersurface
$\Sigma_{1}$ 
 and a final Cauchy hypersurface 
$\Sigma_{2}$. Associated with every such
pair of instants of time there are maps $e_{1}$ and $e_{2}$
from 
$\Gamma$ to $\cal S$ and there is an isomorphism
\begin{equation}
{\cal T}_{12}\colon{\cal S}\to{\cal S}
\end{equation}
given by
\begin{equation}
{\cal T}_{12}=e_{1}\circ e_{2}^{-1}.
\label{dynamics}
\end{equation}
Given a solution to (\ref{eom}), {\it i.e.}, a point in
$\cal S$, 
the linear map ${\cal T}_{12}$ takes its Cauchy data on
$\Sigma_{2}$ and 
yields the solution that has this data on $\Sigma_{1}$. 
This mapping can be viewed as 
``time evolution'' from $\Sigma_{1}$ to $\Sigma_{2}$ as
represented 
on the space of solutions. To see what this time evolution
means in 
terms of evolution of Cauchy data, let 
us use the isomorphism $e_{1}$, associated with the 
initial surface to identify $\Gamma$ and ${\cal S}$. Using
this 
identification, the mapping 
${\cal T}_{12}$ can be viewed as an isomorphism
$\tau_{12}\colon\Gamma\to
\Gamma$ given by
\begin{equation}
\tau_{12}:=e_{1}^{-1}\circ{\cal T}_{12}\circ
e_{1}=e_{2}^{-1}
\circ e_{1}.
\end{equation}
The map $\tau_{12}$ defines dynamical evolution on $\Gamma$
by 
taking data on the initial surface, evolving it into a
solution of 
(\ref{eom}), and then yielding the new data on the final
surface.

Whether viewed as a map on ${\cal S}$ or on $\Gamma$, time
evolution 
preserves the respective symplectic structure:
\begin{equation}
{\cal T}_{12}^{*}\Omega=\Omega,\quad
\tau_{12}^{*}\omega=\omega.
\end{equation}
This is, of course, just the familiar result for
Hamiltonian systems 
that ``time evolution is a 
canonical transformation''.  The problem of quantization on
curved 
surfaces can be viewed as the problem of transporting the
classical 
dynamical structure just described to the quantum
description of the field 
theory.

Wald gives a very general prescription for constructing a
Fock space quantization 
of a linear field in a globally hyperbolic spacetime
\cite{Wald}.  The key 
step is the construction of the one-particle Hilbert space,
from 
which the Fock space ${\cal F}$ is constructed in the usual
way.  The 
possible choices of one-particle Hilbert space correspond
to choices 
of a (suitable) inner product on the symplectic space 
$({\cal S},\Omega)$ (or $(\Gamma,\omega)$). Given an
 inner product on $\cal S$, one obtains 
the Fock space $\cal F$ and (densely defined) 
field operators $\Phi(\varphi)$, labeled by elements of
${\cal S}$, 
satisfying the CCR algebra
\begin{equation}
[\Phi(\varphi_{1}),\Phi(\varphi_{2})]=-i\Omega(\varphi_{1},
\varphi_{2}) .
\end{equation}
Dynamical evolution from $\Sigma_{1}$ to 
$\Sigma_{2}$ in the Heisenberg picture corresponds to the 
algebra automorphism
\begin{equation}
{\cal T}_{12}\cdot \Phi(\phi):=\Phi({\cal
T}_{12}\cdot\varphi).
\end{equation}
Normally, one expects that dynamical evolution is
implemented by a unitary 
transformation $U_{12}\colon {\cal F}\to {\cal F}$, such
that
\begin{equation}
U_{12}^{-1}\Phi U_{12}={\cal T}_{12}\cdot \Phi.
\end{equation}
Assuming this is the case (but see below), 
given a state $|\psi>$ prepared at the 
time defined by $\Sigma_{1}$ we can define the state at
time 
$\Sigma_{2}$ (in the Schrodinger picture) as
$U_{12}|\psi>$. The 
Tomonaga-Schwinger equation (\ref{TSeqn}), at least
formally,  
describes the change in $U_{12}|\psi>$ as the surface
$\Sigma_{2}$ is 
deformed in spacetime.  One can therefore use $U_{12}$ to
define the 
``physical states'' in the Dirac constraint quantization of 
parametrized field theory. In detail, fix once and for all 
an initial surface $\Sigma_{0}$.  Choose a state
$|\psi_{0}>$ 
in the Fock space and  
apply the unitary transformation corresponding to evolution
from 
$\Sigma_{0}$
to an arbitrary final surface $\Sigma$.  The result can be 
viewed as a $\Sigma$-dependent state satisfying (formally)
the 
constraints (\ref{TSeqn}).  This ``physical state'' is
determined by 
its value $|\psi_{0}>$ on the initial surface, 
and all physical states arise by varying 
the choice of $|\psi_{0}>$.  Of course, what I have
described is just 
a ``many-fingered time'' generalization of the standard
approach to 
solving the Schr\" odinger equation.
Insofar as parametrized field theories 
accurately model already parametrized theories such as
general 
relativity, we can interpret the Wheeler-DeWitt equation as
a 
Schr\" odinger equation in disguise and try to find the
physical 
quantum states using the approach just described.

Unfortunately,  unitary implementability of the symplectic 
transformation ${\cal T}_{12}$ is not guaranteed.  
It is well-known that not all canonical 
transformations can be unitarily implemented in quantum
mechanics. In 
quantum field theory the presence of an infinite number of
degrees of 
freedom can even prevent unitary implementability of linear
canonical 
transformations, such as we are considering here
\cite{Shale}.  
Failure of unitary 
implementability of time evolution is not unheard of in the
context 
of quantum field theory in a non-stationary curved
spacetime 
\cite{Wald}, where it is usually associated with infinite
particle 
production by the gravitational field. However, 
there the whole 
issue is complicated by the fact that the fields are
interacting with 
a prescribed gravitational field and, in general, there is
no preferred 
quantization of the classical theory.  
A much simpler situation one 
might consider is that of free fields in flat spacetime;
there is then 
a 
timelike Killing vector field that defines a 
preferred inner product on ${\cal S}$ and a consequent
preferred 
quantization. Of course 
this is precisely the setting one finds oneself in when
considering 
the simplest of the midi-superspace models described in the
previous 
section.  Following the lead of Kucha\v r \cite{KK89}, in
\cite{TV98} 
M. Varadarajan and I consider unitary 
implementability of dynamical evolution along arbitrary
foliations, as well as the definition and 
solution of the quantum constraints 
(\ref{TSeqn}), for a free scalar field propagating on a
flat two-dimensional 
spacetime.  Let me now summarize some of the salient
results.

The quantum theory for
a free, massless scalar field on a 
flat, cylindrical spacetime
is defined in the 
standard way. Unitary implementability of the 
transformation ${\cal T}_{12}$ can be verified  for 
$\Sigma_{1}$ and 
$\Sigma_{2}$ being any Cauchy surfaces (actually, Cauchy
circles).  
It is possible to construct the unitary transformation
explicitly and 
check the status of the putative quantum constraint
(\ref{TSeqn}). 
Let $T\colon S^{1}\to {\bf R}\times S^{1}$ be an embedding
of a Cauchy 
surface.  We consider the image, $|\Psi(T)>$, of the
unitary map taking any initial 
state on any initial Cauchy surface to the surface embedded
by $T$. We 
find that this state vector satisfies a quantum constraint
of the form
\cite{TV98}
\begin{equation}
\left(
{1\over i}{\delta\over\delta T^{\alpha}} + h_{\alpha} +
A_{\alpha}[T]\right)
|\Psi(T)> = 0.
\end{equation}
Here $h_{\alpha}$ is the normal-ordered (with respect to
the usual 
vacuum) energy-momentum current density in the Schrodinger
picture. 
The term $A_{\alpha}$ is a multiple of the identity
operator which depends 
on the embedding $T$; thus this term is a ``time-dependent
$c$-number''. 
 Based upon general arguments that take into 
account the Schwinger terms in the algebra of
energy-momentum tensor 
components, Kucha\v r proposed that the quantum constraints
for this 
model should have such a term, and our direct computation
verifies this 
proposal.  Thus there is a ``quantum correction'' to the
classical 
constraints, which is an interesting phenomenon to
encounter in such a 
simple model.   

Other, related models one can consider are obtained by
adding a mass 
to the scalar field and/or changing the topology of the
spacetime 
to ${\bf R}^{2}$. Adding a mass does not alter the unitary 
implementability of the ``many-fingered time'' dynamical
evolution. 
Allowing the Cauchy surfaces to be non-compact requires 
asymptotically (extrinsically) flat Cauchy surfaces to be
used in 
order to guarantee unitary evolution in the massive case. 
Massless 
fields on ${\bf R}^{2}$ lead to the usual infrared
difficulties, and 
haven't been explored as yet. Presumably, these results can
be 
generalized to any free fields in two spacetime dimensions. 
Aside from the technical difficulty with massless fields on
${\bf R}^{2}$, 
it appears that the problem of quantization on curved
surfaces is 
satisfactorily solved for free fields in two spacetime
dimensions.  

It is now tempting to suppose that these results have a 
straightforward generalization to free field 
theories on higher-dimensional flat spacetimes.  However,
in quantum field theory it seems that 
nothing should be taken for granted.  If $\Sigma_{1}$
and/or 
$\Sigma_{2}$ are suitably generic, the symplectic
transformation 
representing dynamical evolution will {\it not} be
unitarily implementable. 
A sketch of this result is given in \cite{TV98}; details
will appear in the 
near future \cite{TV??}. The 
situation is rather like the Van Hove obstruction to
unitary 
implementability of the group of canonical transformations
\cite{VH}. In 
quantum mechanics it is well known that only a subset of
the canonical 
transformations can be represented as unitary
transformations on a 
Hilbert space. For a free field theory on flat spacetime
one can 
unitarily
implement dynamical evolution between Cauchy surfaces
related by an 
action of the Poincar\' e group, but more general
transformations  
are not supported by the Fock space. The problem seems to
be that the group of 
hypersurface deformations for a free field does not have a 
unitary representation on the standard free field Fock space
except for two-dimensional spacetimes. For two-dimensional
field 
theories, the many-fingered time evolution can be viewed as
an action 
of the conformal group on spacetime \cite{KT89}.  The
conformal group 
for a two-dimensional spacetime is, in 
turn, built from a pair of one-dimensional diffeomorphism
groups, 
which have unitary representations on Fock space
\cite{Segal}, at 
least for free fields. 
In higher-dimensions this 
simple picture is simply not available.  Thus 
Dirac's problem of quantization on curved 
surfaces  remains an open 
problem  for free fields on a flat spacetime of dimension
greater 
than two.  By this I mean that it is not known how to find
a Hilbert 
space representation of the CCR that allows for unitary
dynamical 
evolution along arbitrary foliations and which reduces to
the usual 
dynamical evolution when using foliations by flat
hypersurfaces.  
It is worth noting that there seems to be no obstruction to 
quantization on curved surfaces using the more general
approach to 
quantum theory
provided by the
apparatus of algebraic 
quantum theory (see \cite{Haag,Wald} for a description of
this approach 
to quantum field theory).  

At this point it is appropriate to recall the motivation
for our 
discussion of quantization on curved surfaces, that is,
quantization 
of parametrized field theories.  We were able, classically,
to view 
the two Killing vector midi-superspaces as the
parameterization of field 
theories on a flat three-dimensional spacetime.  It is then
natural 
to try to define 
 the ``Wheeler-DeWitt equation'' as the functional Schr\"
odinger 
equation (\ref{TSeqn}).  We saw that quantization of
parametrized  
fields in dimensions three and higher is problematic; 
what is to be done about the 
quantization of these midi-superspace models {\it a la}
Dirac? 
This question is made 
more interesting by the fact that the reduced field
theories for these 
models, while naturally viewed as symmetry reductions of 
three-dimensional theories, are in effect two-dimensional
theories. Is 
it possible that these models are just two-dimensional
enough in their 
behavior to allow the quantization via parametrized field
theory?  
Recent 
work of M. Varadarajan suggests that this is the case
\cite{MV}. 
Work is in progress on this issue, and I hope that soon we
will know 
whether the quantization of midi-superspace models of
canonical 
quantum gravity
proposed in \cite{Kuchar1971} and \cite{RT96} is possible.
If so, results 
such as \cite{KS,Ashtekar96} can be viewed as arising from
a generally 
covariant canonical 
quantization of the gravitational field, and these models
can shed 
light on the nature of quantum geometry in the framework of
Dirac's 
quantization of constrained systems.

\acknowledgments

The work described here was performed in collaboration with
Ian 
Anderson and Mark Fels (Utah State University), with Joseph
Romano 
(Californian Institute of Technology), and with Madhavan
Varadarajan (Raman Research Institute).
I would like to thank the organizers of Quantum Gravity in
the 
Southern Cone II for inviting me to speak, and for their
warm hospitality. 
This work was supported in 
part by grant PHY-9600616 from the National Science
Foundation.

%
%


\begin{references}
\bibitem{note2} Sometimes also called a ``harmonic map''. 
The 
adjective ``wave''
is more appropriate in this context since
the field 
equations are hyperbolic.
\end{references}
\end{document}